\documentclass[twocolumn,pre,showpacs]{revtex4}
\usepackage{graphicx,color}
\usepackage{dcolumn}
\usepackage{amsmath,amssymb}

\def\KaFST{-1.97}
\def\KaBKT{1.523}
\def\KvBKTKh0{0.621}
\def\discrepancy{0.01\%}
\def\anisotropy{1.5}
\def\figurescale{0.93}
\def\figurescalelarge{0.95}

 \def\HYOUichi{
 \begin{table}[t]
  \caption{
  Discrete symmetries of the principal operators
  ($O_{4,5}$ are referred to in Sec.\ \ref{sec_DISCUSSIONS}).
  The expressions $M$, $k_x$, and $P$ represent the string number,
  the momentum in the $x$ direction, and the parity for the reflection
  $\sigma_x$, respectively (see text). 
  Also given are notations and identifications in both dimer and
  quantum-spin (optional) languages.
  } 
  \begin{tabular}{cccccccc}
   \hline\hline
   Notations & Operators               &&Identifications  &&$M$   &$k_x$&$P$ \\
   \tableline
   $O_0$     &$\sqrt2\cos\sqrt2\phi$   &&HC (N\'eel)      &&0     &$\pi$&$-1$\\
   $O_{1,2}$ &$\exp(\pm i\sqrt2\theta)$&&monomer (doublet)&&$\pm1$&  0  & +1 \\
   $O_3$     &$\sqrt2\sin\sqrt2\phi$   &&VC (dimer)       &&0     &  0  & +1 \\
   $O_4$     &$\sqrt2\cos2\sqrt2\phi$  &&Orientational    &&0     &$ 0 $&$+1$\\
   $O_5$     &$\sqrt2\sin2\sqrt2\phi$  &&Plaquette        &&0     &$\pi$&$-1$\\
   \hline\hline
  \end{tabular}
  \label{tabI}
 \end{table}
 }

 \def\HYOUni{
 \begin{table}[t]
  \caption{
  Estimations of the BKT and the first-order transition points in the
  isotropic system.
  For the BKT transition the rotational order parameters were treated in
  Ref.\
  \cite{Alet05}.
  In Refs.\
  \cite{Alet05,Cast07},
  the first-order transition point were estimated from the breakdown of
  the condition $c=1$.
  } 
  \begin{tabular}{cccccccc}
   \hline\hline
                      && Criteria                  & BKT        &&Criteria& First-order\\
   \tableline
   Ref.\ \cite{Alet05}&& Order parameters          & 1.54       &&$c\ne1$ &   $-2.23$  \\ 
   Ref.\ \cite{Cast07}&& $c$, $x$, etc.            &$1.5-1.7$   &&$c\ne1$ &   $-1.39$  \\
   Present            && Equation\ (\ref{eq_LS_L4})&\KaBKT      &&$K=0$   &  $\KaFST$  \\
   \hline\hline
  \end{tabular}
  \label{tabII}
 \end{table}
 }

 \def\ZUichi{
 \begin{figure}[t]
  \begin{center}
   \includegraphics[width=\figurescale\linewidth]{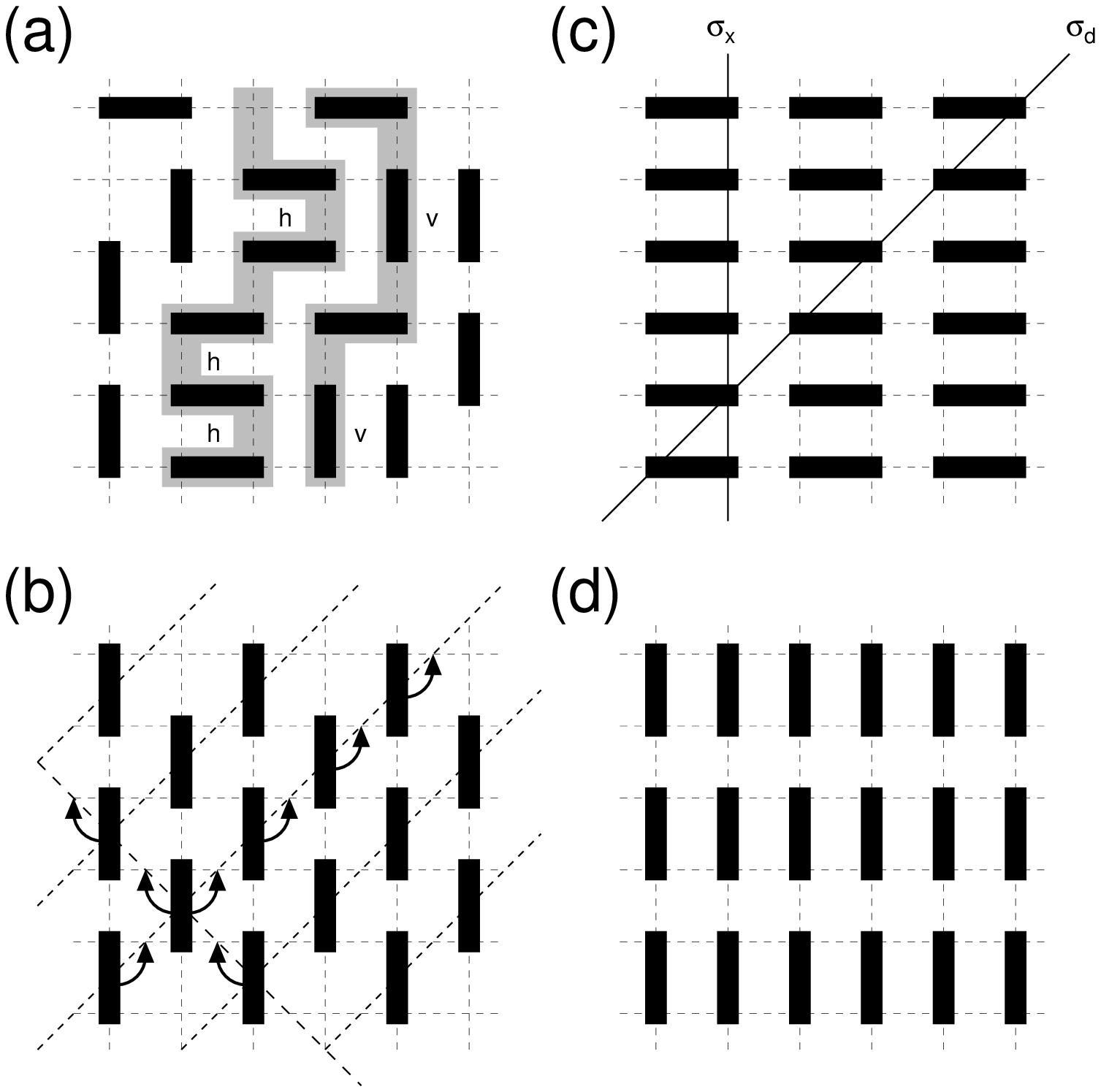}
  \end{center}
  \vspace{-5mm}
  \caption{
  Dimer configurations.
  (a), (b), (c), and (d) represent examples of the liquid, the staggered,
  the HC, and the VC states, respectively.
  The local Boltzmann weights are also given in (a).
  The string representation for (a) using (b) as ``reference state''
  (see text) is given by the gray lines in the $y$ direction.
  In (b), a dotted line in the [11] direction indicates a counterclockwise
  rotation of
  all dimers along the line;
  a dashed line in the $[1\bar{1}]$ direction exhibits a clockwise
  rotation.
  In (c), elements $\sigma_x$ and $\sigma_d$ of the
  ${\bf C}_{4v}$-point group representing reflections about solid lines
  in the $y$ and the diagonal directions are indicated.
  }
  \label{dimerconf}
 \end{figure}
 }

 \def\ZUni{
 \begin{figure}[t]
  \begin{center}
   \includegraphics[width=\figurescale\linewidth]{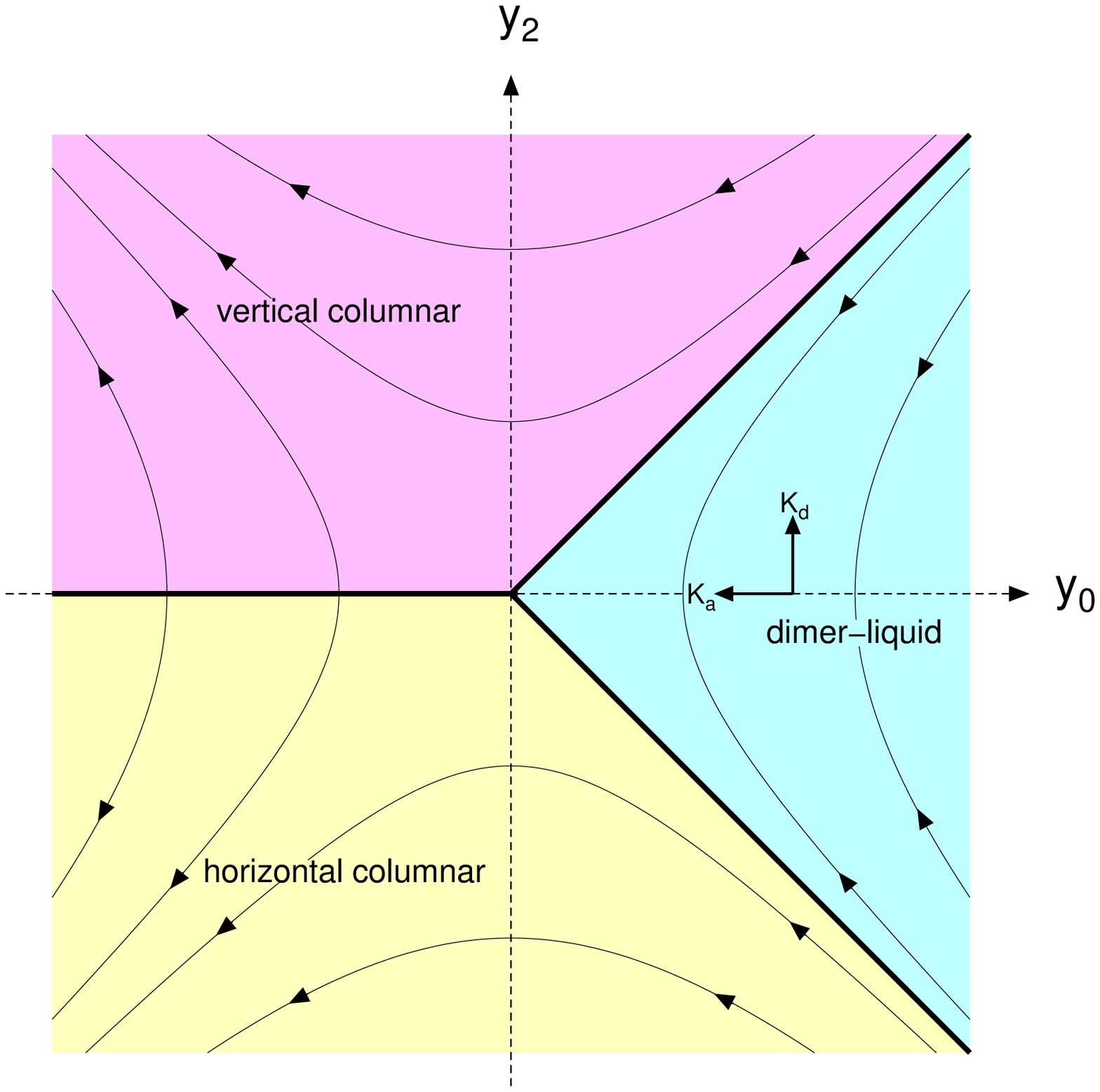}
  \end{center}
  \vspace{-5mm}
  \caption{
  (Color online)
  A schematic representation of the BKT RG-flow diagram around the
  origin of the $(y_0,y_2)$ plane.
  The coordinate frame of the average of couplings $K_a$ and the
  difference of couplings $K_d$ (see text) is present as an inset.
  }
  \label{rgflow}
 \end{figure}
 }

 \def\ZUsan{
 \begin{figure}[t]
  \begin{center}
   \includegraphics[width=\figurescale\linewidth]{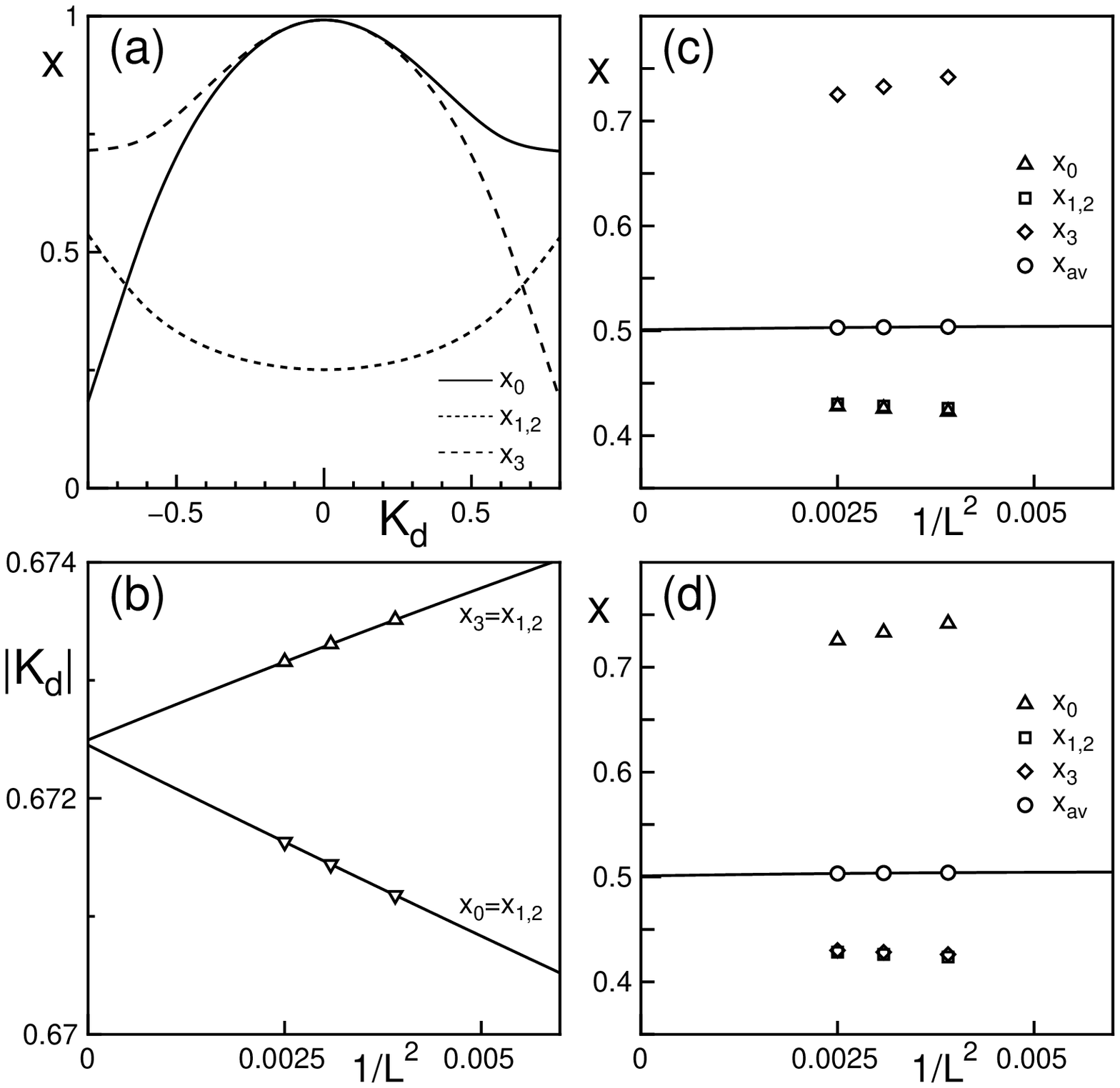}
  \end{center}
  \vspace{-5mm}
  \caption{
  (a)
  An example of 
  the $K_d$ dependences of scaling dimensions at $K_a=0$
  (for $L=20$ case).
  An inserted key identifies excitations and lines.
  (b)
  The finite-size estimates of the transition points to the HC (VC)
  phase denoted by downward (upward) pointing triangles are extrapolated
  according to Eq.\ (\ref{eq_extrap}) (see solid lines).
  (c) and (d)
  Checks of the universal level-splitting conditions\
  (\ref{eq_DLOG_NEEL})
  and
  (\ref{eq_DLOG_DIMER})
  at the BKT-transition points.
  An inserted key identifies excitations and marks;
  $x_{\rm av}$ in
   (c)
  [(d)]
  is the LHS of
   Eq.\ (\ref{eq_DLOG_NEEL})
  [Eq.\ (\ref{eq_DLOG_DIMER})],
  and the least-squares fit solid line is exhibited.
  }
  \label{spector}
 \end{figure}
 }

 \def\ZUyon{
 \begin{figure}[t]
  \begin{center}
   \includegraphics[width=\figurescale\linewidth]{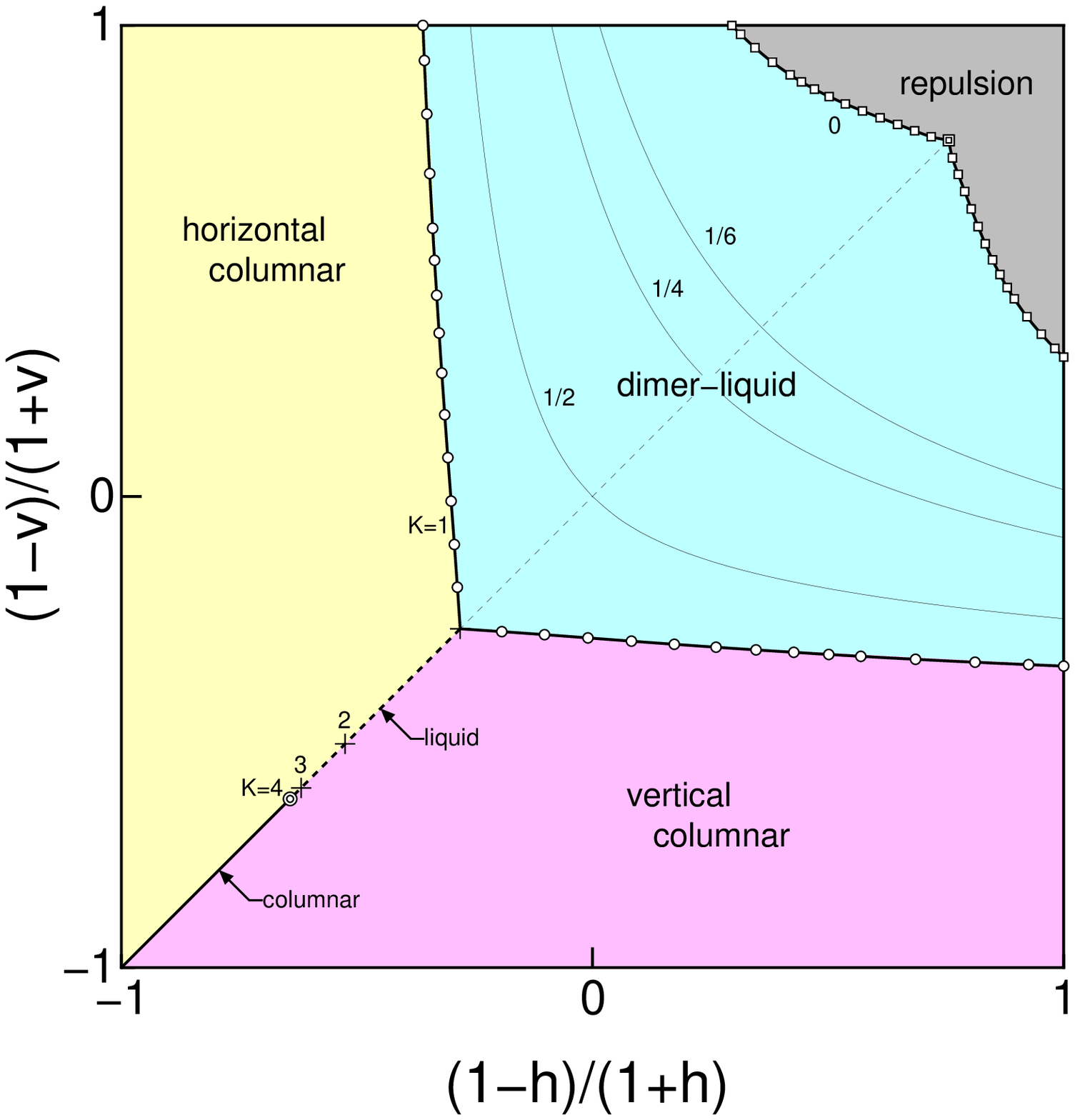}
  \end{center}
  \vspace{-5mm}
  \caption{
  (Color online)
  Global phase diagram.
  The diagonal line (the center point) corresponds to the isotropic
  (non-interacting) system. 
  The circles with solid lines separate the dimer-liquid phase from the
  HC and the VC phases.
  The squares with solid lines exhibit the condition $K=0$, which is the
  boundary of the dimer-liquid phase.
  Also plotted are the contour lines of $K=\frac16$, $\frac14$, and
  $\frac12$ as well as the points (plus marks) of $K=1$, $2$, and $3$.
  The double circle indicates another BKT-transition point brought about
  by ${\cal L}_4$.
  }
  \label{phasediagram}
 \end{figure}
 }

 \def\ZUgo{
 \begin{figure}[t]
  \begin{center}
   \includegraphics[width=\figurescale\linewidth]{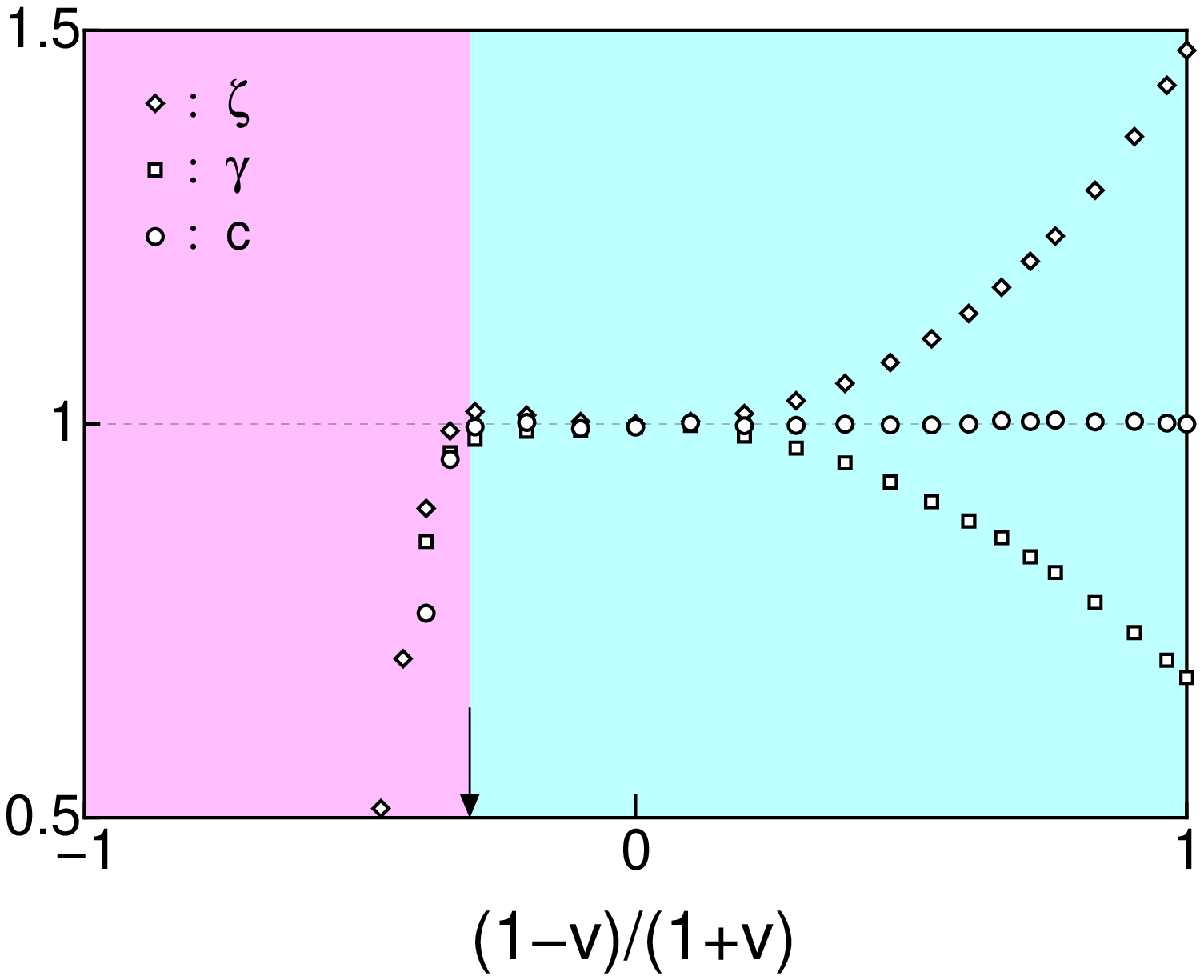}
  \end{center}
  \vspace{-5mm}
  \caption{
  (Color online)
  The $K_v$ dependence of $\zeta$, $\gamma$, and $c$ at $K_h=0$.
  An inset identifying marks and physical quantities is given.
  The phase boundary between the dimer-liquid and the VC phases is
  around $K_v\simeq\KvBKTKh0$ (see the arrow).
  }
  \label{eta-cv-c}
 \end{figure}
 }

 \def\ZUnana{
 \begin{figure}[t]
  \begin{center}
   \includegraphics[width=\figurescale\linewidth]{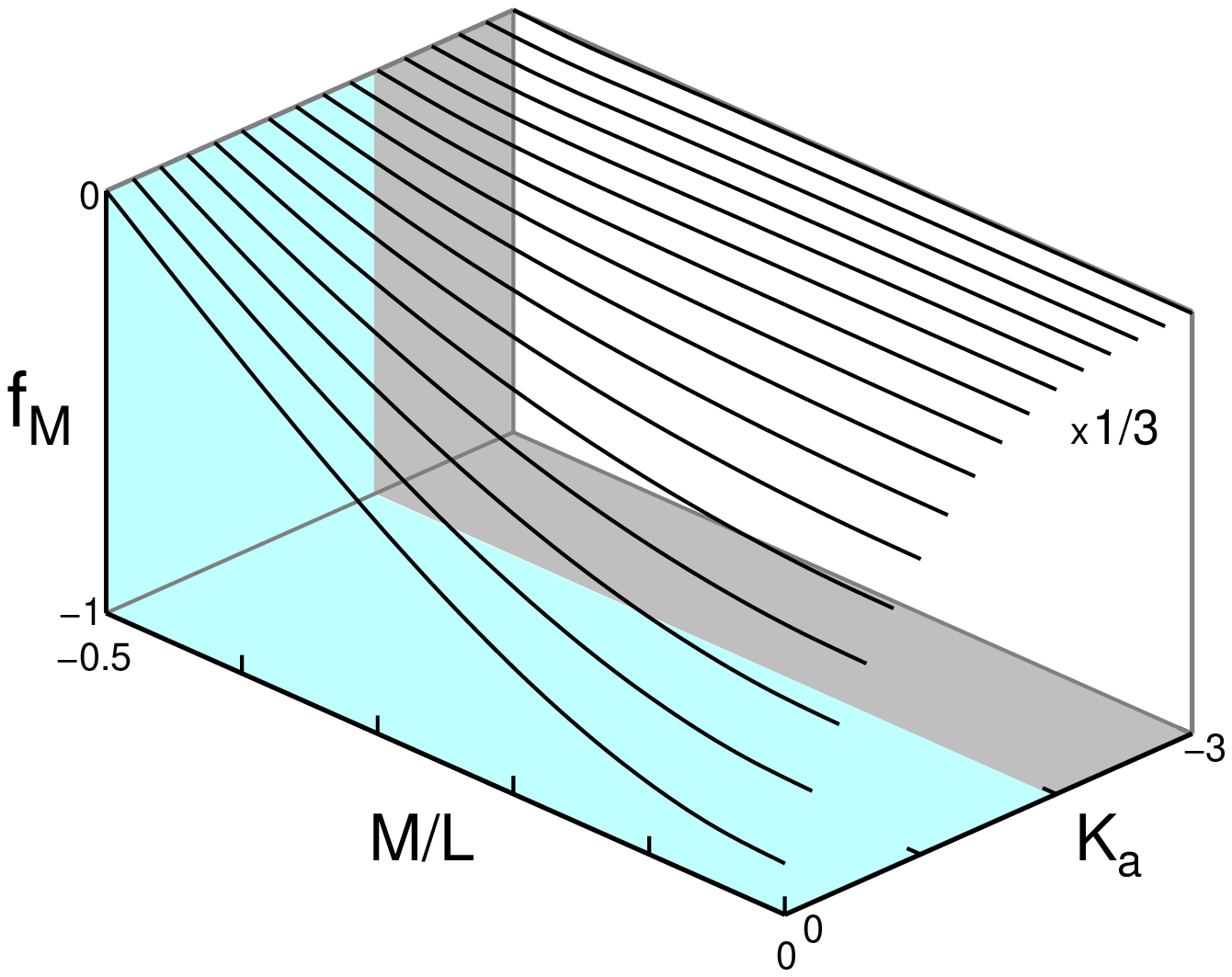}
  \end{center}
  \vspace{-5mm}
  \caption{
  (Color online)
  The free-energy density $f_M$ as a function of the string number $M$
  and the interaction $K_a$ for the isotropic system with $L=20$.
  The phase boundary between the dimer-liquid and the strong repulsion
  phases is estimated as $K_a\simeq\KaFST$.
  }
  \label{freeenergy}
 \end{figure}
 }

 \def\ZUhachi{
 \begin{figure}[t]
  \begin{center}
   \includegraphics[width=\figurescale\linewidth]{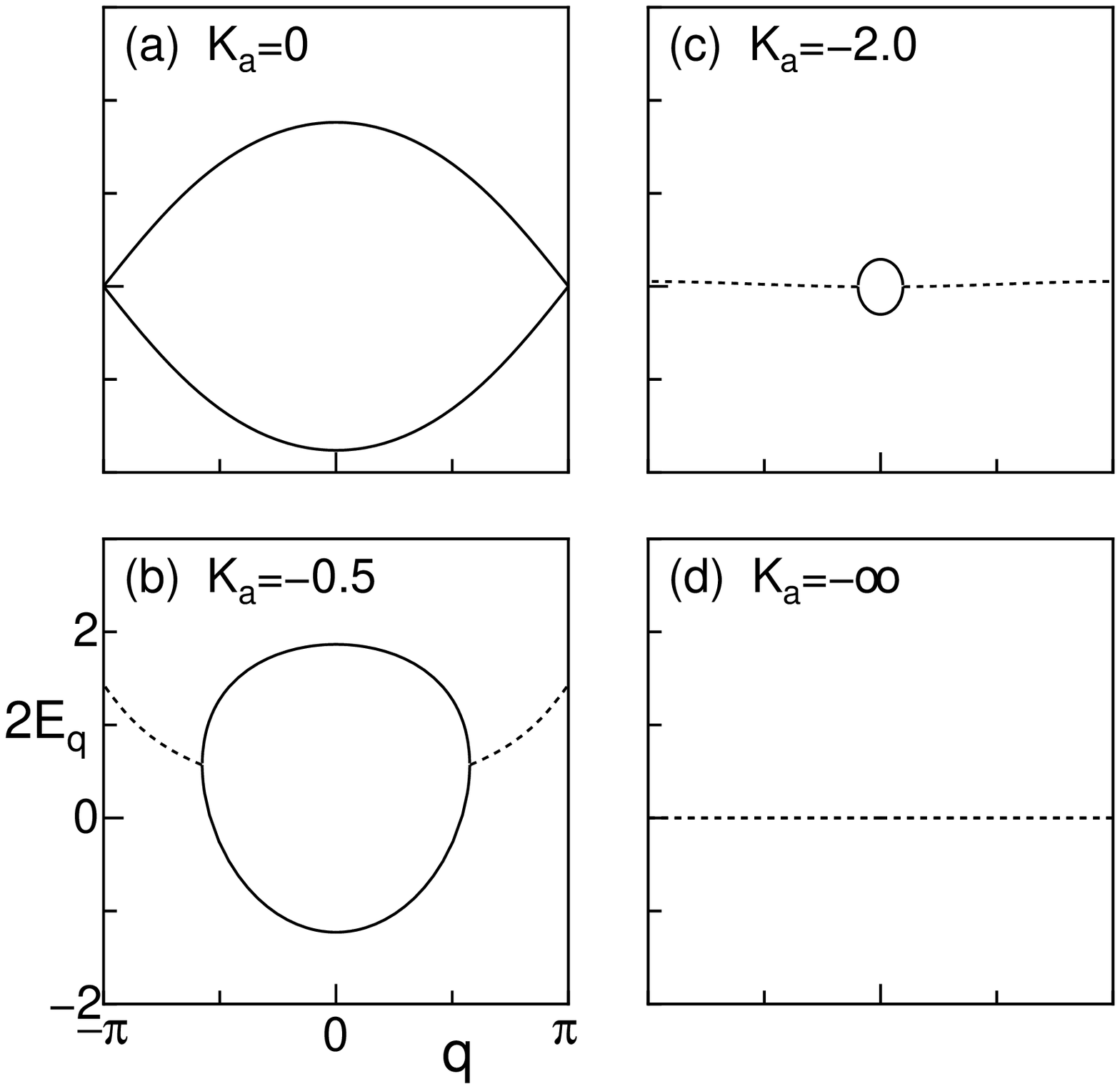}
  \end{center}
  \vspace{-5mm}
  \caption{
  Analytical results for the dispersion relations of a one-string
  motion.
  The lower two of the four bands are drawn for the isotropic systems.
  The complex values are two-fold degenerate and denoted by dotted
  lines; the flat band appears in the strong repulsion limit [panel (d)].
  }
  \label{oneparticle}
 \end{figure}
 }

 \def\ZUqu{
 \begin{widetext}
 \begin{figure*}[t]
  \begin{center}
   \includegraphics[width=\figurescalelarge\linewidth]{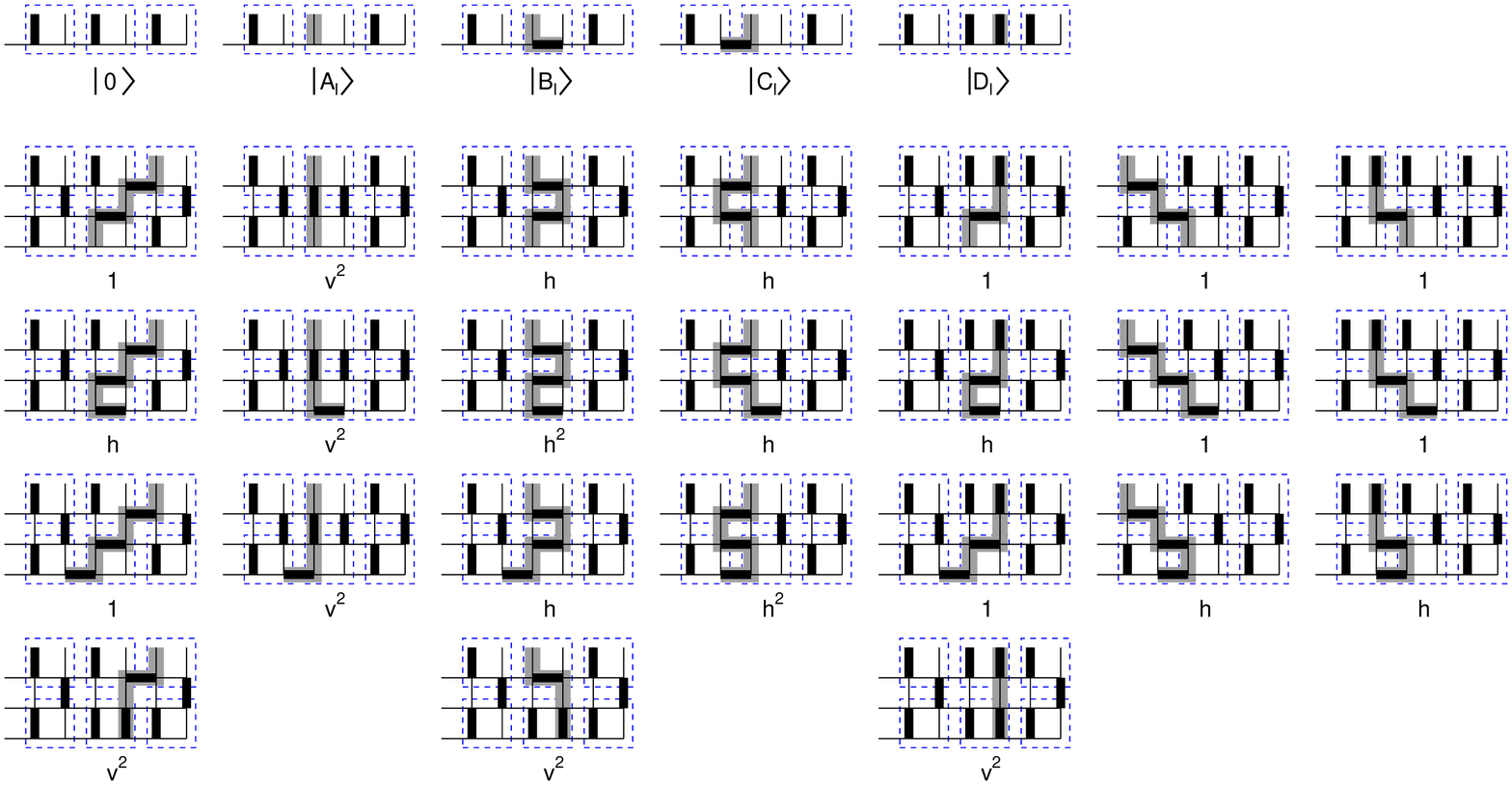}
  \end{center}
  \vspace{-5mm}
  \caption{
  (Color online)
  The string vacuum
  $\left|0\right\rangle$
  and the one-string states
  $\{\left|A_l\right\rangle$,
  $\left|B_l\right\rangle$,
  $\left|C_l\right\rangle$,
  $\left|D_l\right\rangle\}$
  are depicted in the first line.
  A square given by a dotted blue line indicates a unit cell which
  includes two sites and four bonds.
  Dimers and strings are given by black rectangles and gray lines,
  respectively.
  In subsequent lines,
  24 microscopic processes of transfers of one-string states between two
  next-nearest-neighbor rows are given with weights.
  }
  \label{states-elements}
 \end{figure*}
 \end{widetext}
 }

\begin{document}

 \title{
 Classical dimer model with anisotropic interactions on the square
 lattice
 }
 \author{
 Hiromi Otsuka
 }
 \address{
 Department of Physics, Tokyo Metropolitan University, Tokyo 192-0397,
 Japan
 }
 
 \date{\today}

\begin{abstract}

 We discuss phase transitions and the phase diagram of a classical dimer
 model with anisotropic interactions defined on a square lattice.
 For the attractive region,
 the perturbation of the orientational order parameter introduced by the
 anisotropy causes the Berezinskii-Kosterlitz-Thouless transitions from
 a dimer-liquid to columnar phases.
 According to the discussion by Nomura and Okamoto for a quantum-spin
 chain system
 [J. Phys. A {\bf 27}, 5773 (1994)],
 we proffer criteria to determine transition points and also
 universal level-splitting conditions.
 Subsequently, we perform numerical diagonalization calculations of the
 nonsymmetric real transfer matrices up to linear dimension specified by
 $L=20$ and determine the global phase diagram.
 For the repulsive region, 
 we find the boundary between the dimer-liquid and the strong repulsion
 phases.
 Based on the dispersion relation of the one-string motion, which
 exhibits a two-fold ``zero-energy flat band'' in the strong repulsion
 limit, we give an intuitive account for the property of the strong
 repulsion phase.

 \end{abstract}
 \maketitle

 \section{INTRODUCTION}
 \label{sec_INTRO}

 In the early 1960s, Kasteleyn
 \cite{Kast61}
 and
 Temperley and Fisher
 \cite{Temp61,Fish61}
 studied the classical dimer model (DM) defining the
 statistical-mechanical problem of the covering of a lattice
 by dimers.
 They treated the DM on, for example, the square lattice in
 the thermodynamic limit, and obtained the partition function
 to give
 the extensive entropy of an ensemble of dimer configurations.
 In particular, it was shown that the close-packed model defined on
 planar lattices can be solved exactly by Pfaffian techniques
 \cite{Kast63}.
 The properties of these ensembles were studied in subsequent research.
 For example, those defined on bipartite (non-bipartite) lattices
 such as the square (triangular) lattice exhibit critical (off-critical)
 behavior
 \cite{Fish-Step63,Fend02},
 which are now thought to reflect an existence (absence) of the height
 representations for the DMs
 \cite{Blot82,Blot93,Henl97,Ardo04,Kond95Kond96}.
 While the relevance of dimers as diatomic molecules adsorbed on a
 lattice is clear and direct, it can be also related to other degrees of
 freedom
 \cite{Fish66}.
 For instance, the zero-temperature Ising-spin antiferromagnet on a
 triangular (Villain) lattice can be related to the DM on a hexagonal
 (square) lattice, where each dimer represents
 an unsatisfied bond of spins
 \cite{Blot82,Blot93,Vill77}.
 Also widely known is the string representation whereby
 dimer systems under a certain condition can be related to loop gases
 whose configurations are classified by winding numbers (see below)
 \cite{Blot82,Blot93,Otsu06a}.
 This correspondence has been utilized in discussions of polymer
 systems
 \cite{Jian07}.
 More importantly, Rokhsar and Kivelson introduced quantum dimer
 models (QDMs) to describe the valence-bond physics in quantum
 Heisenberg antiferromagnets, where the dimer represents a tightly
 binding singlet pair of quantum spins
 \cite{RK}.

 These are but a fraction of the examples that show DMs having
 importance in a wide range of research and
 drawing attention over the years;
 in particular, current interest has been mainly focused on exotic
 phases with topological orders observed in the QDMs
 \cite{Moes01}.
 More recently, Blunt et al. reported on an adsorption
 experiment of certain rod-like organic molecules on graphite
 \cite{Blun08}
 and explained its relevance to the DM on a hexagonal lattice which
 includes interactions between neighboring dimers
 \cite{Jaco09}. 
 This exhibits the fact that interaction effects in classical dimers are
 also important from both theoretical and experimental viewpoints
 \cite{Alet05}.

 In this paper, we investigate an interacting dimer model (IDM) defined
 on a square lattice:
 suppose the lattice constant $a=1$ and let $\Lambda$ denote the set of
 lattice sites.
 Then the following reduced Hamiltonian expresses interactions between
 two nearest-neighbor dimers:
 \begin{align}
  {\cal H}
   =-\sum_{(k,l)\in\Lambda}
   \Bigl[&K_h n(k+\frac12,l) n(k+\frac12,l+1)\nonumber\\
   +     &K_v n(k,l+\frac12) n(k+1,l+\frac12)\Bigr],
   \label{eq_Hamil}
 \end{align}
 where lattice sites and lattice bonds are denoted, respectively, as
 $(k,l)$ and $(k+\frac12,l)$ with $k,l\in\mathbb{Z}$.
 We locate the dimer occupation numbers
 $n(k+\frac12,l)=0$ or $1$ (binary variables) on the bonds.
 The first (second) term on the right-hand side of
 Eq.\ (\ref{eq_Hamil}) represents an interaction between parallel
 horizontal (vertical) dimers [see Fig.\ \ref{dimerconf}(a)].
 Defining the local Boltzmann weights as
 \begin{align}
  h=\exp(K_h),~~~ 
  v=\exp(K_v),
  \label{eq_wv}
 \end{align}
 with subscripts on the $K$'s referring to the corresponding dimer pair,
 the partition function $Z(h,v)$ is then expressible as a summation with
 respect to the dimer configurations $C$ on $\Lambda$,
 \begin{align}
  Z(h,v)
  =\sum_C h^{N_h(C)} v^{N_v(C)},
  \label{eq_Z}
 \end{align}
 where $N_h(C)$ [$N_v(C)$] represents the number of plaquettes with 
 parallel horizontal (vertical) dimers.
 For large $h$ or $v$, an attractive case,
 a twofold- or a fourfold-degenerate state with columnar order is
 expected to be stabilized
 [see examples in Figs.\ \ref{dimerconf}(c) and \ref{dimerconf}(d)].
 Meanwhile,
 for small $h$ and $v$, a repulsive case,
 a highly degenerate phase is stabilized
 [see an example in Fig.\ \ref{dimerconf}(b)].
 For the isotropic case $K_h=K_v$, results of some numerical calculations
 are already available in the literature
 \cite{Alet05,Cast07},
 but its extension to the anisotropic case $K_h\ne K_v$ is still lacking.
 Therefore, we shall clarify the global phase diagram and provide
 evidence corroborating the properties of the phase transitions observed in
 anisotropically interacting dimers.

 \ZUichi

 According to the effective field theory discussed by
 Papanikolaou, Luijten, and Fradkin (PLF)
 \cite{Papa07},
 one effect absent in the isotropic system is a perturbation by
 the orientational order parameter; this brings about
 Berezinskii-Kosterlitz-Thouless (BKT) transitions
 \cite{Bere71,Kost73Kost74}.
 Another effect is a renormalization of the so-called geometric factor,
 which becomes important in numerical calculations of universal
 quantities, e.g., the central charge and the scaling dimensions of
 operators (see below).
 In such cases, as demonstrated in our own research on an
 antiferromagnetic Potts model with anisotropic next-nearest-neighbor
 couplings, the so-called level-spectroscopy analysis
 \cite{Nomu94,Nomu95}
 can provide an effective way to determine phase transition points
 \cite{Otsu06b}.
 For this reason, we shall also employ the same strategy for the present
 model
 \cite{Otsu05Otsu08}.

 For later convenience,  we shall briefly explain here the string
 representation of the DM on $\Lambda$.
 As explicitly explained in Ref.\
 \cite{Bogn04},
 the transformation of a dimer configuration, e.g., 
 Fig.\ \ref{dimerconf}(a),
 to a string configuration is performed via an
 \texttt{XOR}
 operation with
 reference configuration
 [Fig.\ \ref{dimerconf}(b)].
 The
 \texttt{XOR}
 operation takes the exclusive
 \texttt{OR}
 between occupation (binary)
 numbers in these two configurations over each bond.
 Consequently, we obtain strings running in the $y$ direction
 [see the two gray lines in Fig.\ \ref{dimerconf}(a)].
 Due to the close-packing condition, they have no end points, and thus
 the string configurations for a $L\times L$ system
 ($L$ is an even number)
 with periodic boundary conditions can be characterized by winding
 numbers $(N_x,N_y)$ satisfying $0\le N_{x,y}\le L$.
 While in the numerical calculation of the transfer matrices we shall
 employ $N_y$ as a conserved quantity in the row-to-row transfer of
 configurations, we would rather use a quantity
 \begin{align}
  M\equiv N_y-L/2
 \end{align}
 $(|M|\le L/2)$ for convenience in our discussion.

 The organization of this paper is as follows: 
 In Sec.\ \ref{sec_THEORY}, 
 we review previous research results to give an effective
 description of the low-energy and long-distance behavior of the IDMs
 \cite{Papa07}.
 In particular, the operator content of the theory, including expressions of
 local order parameters and defect operators, and their scaling
 dimensions are explained in detail.
 We then provide the conditions to determine the BKT-transition points
 and clarify some universal relations among excitation levels, which
 serve as a check of our calculational results.
 In doing this, a correspondence with a frustrated quantum-spin chain
 system plays a guiding role.
 Thus, this correspondence will be emphasized and referred to when
 appropriate.
 In Sec.\ \ref{sec_NUMERICAL},
 we summarize our numerical study and results obtained by the
 transfer-matrix calculations based on the conformal field theory (CFT)
 \cite{BPZ}.
 First, we demonstrate that the theoretical predictions in
 Sec.\ \ref{sec_THEORY} can be observed precisely via numerical
 analysis of the excitation spectra.
 Next, we provide the global phase diagram of interacting dimers, which
 includes the BKT, the second-order, and the first-order transition
 lines.
 Also, in a strong repulsion region, we expect a highly degenerate
 phase including the staggered state.
 We calculate the string-number dependence of the free-energy density
 for the isotropic case.
 Furthermore, we investigate the ``dispersion relation'' of the one-string
 motion, and then based on these data we shall try to give an insight 
 into properties of the strong repulsion phase.
 The last section,
 Sec.\ \ref{sec_DISCUSSIONS},
 is devoted to discussion and summary. 
 We also provide our method to evaluate the original BKT transition in
 the isotropic system.
 Finally, we compare our data with previous research results.

 \section{THEORY}
 \label{sec_THEORY}

 Continuum field theories offer unified approaches to investigate
 phase transitions in interacting systems on lattices.
 They are derived in the scaling limit, $a\to0$ while keeping
 ${\bf x}=(x_1,x_2)=(a k, a l/\zeta)$ finite.
 Here $\zeta$ is the geometric factor, taking a fixed value, e.g.,
 $\zeta=1$ $(2/\sqrt3)$ for isotropic systems on a square (triangular)
 lattice.
 However, for anisotropic systems, renormalization of $\zeta$ is
 necessary due to interactions leading to non-universal values.
 The renormalized $\zeta$ can be also related to the velocity of an
 elementary excitation observed in Tomonaga-Luttinger liquids
 \cite{Hald80}.
 Thus, $\zeta$ disappears from the theoretical description if we
 properly employ its renormalized value; but as we will see in
 Sec.\ \ref{sec_NUMERICAL},
 it becomes rather important in numerical calculations.

 According to PLF,
 the effective description of the IDM takes the form of a sine-Gordon
 field theory
 \cite{Alet05}.
 In the present case,
 its expression is 
 given by the Lagrangian density
 ${\cal L}={\cal L}_0+{\cal L}_2+{\cal L}_4$
 with 
 \begin{align}
  &{\cal L}_0
  =
  \frac{K}{2\pi}
  \left(\nabla{\phi}\right)^2,
  \label{eq_L0}\\
  &{\cal L}_2
  =
  \frac{y_2}{2\pi\alpha^2}:\cos2\sqrt2\phi:,
  \label{eq_L2}\\
  &{\cal L}_4
  =
  \frac{y_4}{2\pi\alpha^2}:\cos4\sqrt2\phi:.
  \label{eq_L4}
 \end{align}
 We denote the course-grained height field in the two-dimensional
 Euclidean space as
 $\phi({\bf x})$,
 which satisfies a periodicity in height space of
 $\sqrt2\phi=\sqrt2\phi+2\pi N$
 ($N\in\mathbb{Z}$)
 \cite{Henl97}.
 Due to the close-packing of the dimers, the defect operators given in
 terms of the disorder field $\theta$ dual to $\phi$ are absent from
 ${\cal L}$, so that it represents a roughening phase or flat phases of
 an interface model in three dimensions.
 The theoretical parameter $K$ (the Gaussian coupling) describes the
 stiffness, and determines the dimensions of the operators on the
 Gaussian fixed line ${\cal L}_0$.
 In our notation, the vertex operator with $m$ electric and $n$
 magnetic charges is given by
 ${\cal O}_{m,n}=e^{im\sqrt2\phi+in\sqrt2\theta}$
 whose dimension is
 \begin{align}
  X_{m,n}=\frac12\left(K^{-1}m^2+Kn^2\right).
 \end{align}
 Therefore, $K=1$ ($K=4$) represents the condition that the perturbation
 ${\cal L}_2$ (${\cal L}_4$) becomes marginal.

 \ZUni

 To make our discussion more concrete,
 we introduce the average ($a$) and the difference ($d$) of the
 couplings as
 \begin{align}
 K_{a,d}=\frac12\left(K_v\pm K_h\right)
 \end{align}
 where the first (second) subscript refers to the upper (lower) sign.
 Then,
 around the non-interacting point, the parameters in ${\cal L}$
 are roughly given by
 \begin{align}
  K\simeq\frac12+c_1K_a,~~
  y_2\simeq c_2K_d,~~
  {\rm and~~}
  y_4\simeq-c_3
 \end{align}
 ($c_{1,2,3}>0$).
 We see that the attractive interaction $K_a>0$ increases $K$, and tends
 to stabilize the columnar states.
 Also, $y_2$ in ${\cal L}_2$ (the orientational order parameter) is
 proportional to the difference, $K_d$, while $y_4$ in ${\cal L}_4$
 which is a remnant from the discreteness of the square lattice is
 almost constant.
 For $K>4$, both nonlinear terms are relevant, but they are not
 competing against each other, so the fourfold-degenerate columnar
 state stabilized by ${\cal L}_4$ is only lifted to realize
 twofold-degenerate columnar states by ${\cal L}_2$ (see below).
 Since the BKT transition by ${\cal L}_4$ was already discussed in the
 literature
 \cite{Alet05,Cast07},
 we shall focus our attention on the role of ${\cal L}_2$.

 According to the standard argument
 \cite{Poly72},
 the renormalization-group (RG) flow diagram of the sine-Gordon model
 ${\cal L}_0+{\cal L}_2$ (here ${\cal L}_4$ is irrelevant) is expressed
 by the BKT RG equations
 \cite{Bere71,Kost73Kost74};
 we depict it by employing the coupling constants $y_0$ $(=2-2K)$ and
 $y_2$ in Fig.\ \ref{rgflow}, where the separatrices $y_2=\mp y_0$
 separate the dimer-liquid phase from two types of twofold-degenerate
 columnar phases,
 namely, 
 the horizontal columnar (HC) state consisting of dimers in the
 horizontal direction
 and 
 the vertical columnar (VC) state consisting of dimers in the vertical
 direction.
 Now, we can point out that our task to treat the BKT transitions in
 the IDM can be related to the investigation of the spin-$\frac12$
 XXZ chain with next-nearest-neighbor interaction because these share
 the same effective description
 \cite{Nomu94,Nomu95}.
 To specify the relationship, we introduce the following operators:
 \begin{align}
  O_0&=\sqrt2\cos\sqrt2\phi,
  \label{eq_Neel}\\
  O_{1,2}&={\rm exp}(\pm i\sqrt2\theta),
  \label{eq_Double}\\
  O_3&=\sqrt2\sin\sqrt2\phi.
  \label{eq_Dimer}
 \end{align}
 Here $O_{0,3}$ stand for the horizontal and the vertical components of
 the columnar local order parameter, and take expectation values
  $\langle O_0\rangle\ne0$ and $\langle O_3\rangle = 0$
 ($\langle O_0\rangle = 0$ and $\langle O_3\rangle\ne0$)
 in the HC (VC) phase;
 $O_{1,2}$ are the defect (or monomer) operators which change the
 winding numbers classifying dimer configurations.
 Alternatively, in the quantum-spin chain language,
 Eqs.\ (\ref{eq_Neel})--(\ref{eq_Dimer})
 correspond to
 the N\'eel,
 the doublet, and
 the dimer operators,
 and give the lowest excitations in
 the N\'eel,
 the spin-liquid, and
 the dimer phases,
 respectively (see Table\ \ref{tabI}).
 Nomura and Okamoto (NO)
 provided criteria to determine the BKT-transition points between
 the spin-liquid and the N\'eel (or dimer) phases based on one-loop
 calculations of the scaling dimensions of these operators
 \cite{Giam89}.
 Therefore, following their argument, we shall discuss procedures to
 determine the BKT-transition points in our IDM.

 \HYOUichi

 Consider a system with a finite-strip geometry, viz., a narrow band of
 width $L$ along the $x$ direction and infinite length along the $y$
 direction.
 The periodic boundary condition is imposed across the width of the
 strip.
 The finite-size corrections to the scaling dimensions of the above
 operators are our key quantities to be evaluated analytically and
 numerically.
 Here, we first consider the system near the separatrix $y_2=-y_0$ where
 a small parameter $t$ can be introduced, so that $y_2=-y_0(1+t)$
 ($|t|\ll1$).
 Next, the conformal perturbation calculations of the renormalized
 scaling dimensions were performed using the sine-Gordon Lagrangian
 density, for which the results can be summarized as follows:
 \begin{align}
  x_0&\simeq\frac12-\frac14 y_0(l)\left(1+2t\right),
  \label{eq_XNeel_NEEL}\\
  x_{1,2}&\simeq\frac12-\frac14 y_0(l),
  \label{eq_XDouble_NEEL}\\
  x_3&\simeq\frac12+\frac14 y_0(l)\left(3+2t\right),
  \label{eq_XDimer_NEEL}
 \end{align}
 ($l=\ln L$ is the logarithmic scale length)
 \cite{Giam89}.
 According to the discussion by NO, we can find the criterion to
 determine the BKT-transition point $t=0$
 (i.e., the level-crossing condition)
 \begin{align}
  x_0=x_{1,2}
  \label{eq_BKT_NEEL}
 \end{align}
 and the level-splitting condition as
 \begin{align}
  \frac{3x_{0,1,2}+ x_3}{4}=\frac12.
  \label{eq_DLOG_NEEL}
 \end{align}
 The latter is one of the universal relations among the excitation
 levels on the separatrix
 \cite{Zima87},
 and enables us to check the consistency of the calculations.
 Second, we investigate the system near the separatrix $y_2=y_0$;
 it proceeds in an analogous way to the above.
 Writing $y_2=y_0(1+t)$, we then obtain the dimensions as
 \begin{align}
  x_0&\simeq\frac12+\frac14 y_0(l)\left(3+2t\right),
  \label{eq_XNeel_DIMER}\\
  x_{1,2}&\simeq\frac12-\frac14 y_0(l),
  \label{eq_XDouble_DIMER}\\
  x_3&\simeq\frac12-\frac14 y_0(l)\left(1+2t\right).
  \label{eq_XDimer_DIMER}
 \end{align}
 Thus, the level-crossing condition needed to determine the transition
 point is given by
 \begin{align}
  x_{1,2}=x_3, 
  \label{eq_BKT_DIMER}
 \end{align}
 and the level-splitting condition is given by
 \begin{align}
  \frac{x_0+ 3x_{1,2,3}}{4}=\frac12.
  \label{eq_DLOG_DIMER}
 \end{align}
 In analogy to the quantum-spin chain, each level crossing represents an
 emergence of a SU(2) multiplet structure consisting of the singlet
 and the triplet states (e.g., $x_3$ and $x_{0,1,2}$ at $y_2=-y_0$).
 Since these are the low-energy levels in the level-1
 SU(2) Wess-Zumino-Witten model
 \cite{Zima87},
 our criteria,
 Eqs.\ (\ref{eq_BKT_NEEL}) and (\ref{eq_BKT_DIMER}),
 are natural and also convincing from this viewpoint.

 At this stage, two comments are in order about the advantage in using
 these relations (the level-spectroscopy approach) and the structure of
 the phase diagram.
 In Sec.\ III, we will outline the numerical transfer-matrix
 calculations performed to obtain the phase diagram.
 Although it can treat systems with a strip geometry, accessible sizes
 are strongly restricted to small values, e.g., $L\le20$ in our
 calculations.
 In the BKT transition, as seen above, the correction terms are
 typically given by the logarithmic form $y_0\simeq 1/\ln(L/L_0)$.
 If we employ the standard KT criterion such as $x_{0}=\frac12$ to
 determine the transition point,
 its finite-size estimates include these, and thus exhibit a slow
 convergence in their extrapolation to the thermodynamical limit.
 Alternatively, 
 criteria\ (\ref{eq_BKT_NEEL}) and (\ref{eq_BKT_DIMER})
 take the logarithmic corrections into account, so they provide
 finite-size estimates with fast convergences
 \cite{Nomu95}.
 Consequently, we can employ the following least-squares-fitting form in
 extrapolating the finite-size data ${\cal Q}(L)$ to the thermodynamic
 limit $L\to\infty$:
 \begin{align}
  {\cal Q}(L)\simeq {\cal Q}(\infty)+a/L^2+b/L^4,
  \label{eq_extrap}
 \end{align}
 which includes the $1/L^2$ term stemming from the $x=4$ irrelevant
 operators as the leading universal correction
 \cite{Card86a}.

 The biases from the finite-strip geometry disappear in the limit, and
 the phase diagram is symmetric with respect to the isotropic line
 $K_h=K_v$ which is one of the inherent properties of the model.
 Here, we describe how the symmetry of the lattice model is embedded in
 the sine-Gordon field theory. 
 We shall consider the generators of the ${\bf C}_{4v}$-point group:
 the $\pi/2$ rotation ($C_4$) and the reflection in the $x$ axis
 ($\sigma_x$) about the original site
 [see Fig.\ \ref{dimerconf}(c)].
 As well as coordinate transformations, these bring about the following
 changes in the height field
 \cite{Alet05,Papa07}:
 \begin{align}
  C_4: \sqrt2\phi\to\sqrt2\phi-\pi/2,~~\sigma_x: \sqrt2\phi\to\pi-\sqrt2\phi.
  \label{eq_phaseshift}
 \end{align}
 All other elements are obtained from Eq.\ (\ref{eq_phaseshift}); among
 them, we shall investigate transformations of the Lagrangian density
 and the principal operators by reflection about the diagonal line,
 $\sigma_d$ $(=\sigma_x\circ C_4)$ [see Fig.\ \ref{dimerconf}(c)], which
 shifts the field as
 \begin{align}
  \sigma_d:~\sqrt2\phi\to\pi/2-\sqrt2\phi.
 \end{align}
 Since the orientational order parameter ${\cal L}_2$ is odd for
 $\sigma_d$, the Lagrangian density transforms as
 \begin{align}
  \sigma_d:~{\cal L}(K,y_2,y_4)\to{\cal L}(K,-y_2,y_4). 
 \end{align}
 This indicates a connection between the positive and the negative
 values of $K_d$.
 In addition, the above four operators transform as
 \begin{align}
  \sigma_d:~O_0\to O_3,~~O_{1,2}\to O_{1.2},~~O_3\to O_0.
 \end{align}
 Thus,
 the symmetry operation $\sigma_d$ interchanges the roles of the HC and
 the VC operators while leaving unchanged the doublet of the monomer
 excitations.
 Consequently, as expected, the level-crossing and the level-splitting
 conditions for $K_d<0$,
 Eqs.\ (\ref{eq_BKT_NEEL}) and (\ref{eq_DLOG_NEEL}),
 are translated to those for $K_d>0$, 
 Eqs.\ (\ref{eq_BKT_DIMER}) and (\ref{eq_DLOG_DIMER}).
 In Sec.\ \ref{sec_NUMERICAL},
 we shall provide some numerical data to check this symmetry.

 \section{NUMERICAL CALCULATIONS}
 \label{sec_NUMERICAL}

 Now, consider a system on $\Lambda$ with the $L\times\!\infty$ stripe
 geometry and introduce
 the transfer matrix ${\bf T}_M(L)$ connecting nearest-neighbor rows in
 the $y$ direction
 [see Fig.\ \ref{dimerconf}(a)]
 \cite{Alet05,Cast07}.
 As mentioned in Sec.\ \ref{sec_INTRO}, the string number in the $y$
 direction, $M$, is a conserved quantity, which can thus be specified
 explicitly.
 We denote the eigenvalues as
 $\lambda_p(L)$
 and their logarithms as
 $E_p(L)=-\ln|\lambda_p(L)|$
 ($p$ specifies an excitation level such as those listed in the above).
 Then,
 the conformal invariance provides direct expressions of the central
 charge $c$ and the scaling dimension $x_p$ in the critical systems as
 \cite{Blot86,Card84}
 \begin{align}
  E_{\rm g}(L)\simeq Lf-\frac{\pi}{6L\zeta}c,~~~ 
  \Delta E_p(L)\simeq \frac{2\pi}{L\zeta}x_p.
  \label{eq_c_xs}
 \end{align}
 Here,
 $E_{\rm g}(L)$,
 $\Delta E_p(L)$ $[=E_p(L)-E_{\rm g}(L)]$,
 and
 $f$ 
 correspond to
 the ground-state energy,
 an excitation gap,
 and
 a free-energy density,
 respectively.
 The ground state is found in the $M=0$ ($N_y=L/2$) sector, and the
 excited levels are also in the sectors specified by the discrete
 symmetries given in Table\ \ref{tabI}.
 In addition,
 since, independent of the value of $K$, the scaling dimension of a
 level-1 descendant is equal to 1, it has been utilized to estimate a
 velocity of elementary excitation in the Tomonaga-Luttinger liquid
 (see, for example, \cite{Nomu94}).
 According to their treatment, the effective geometric factor (i.e.,
 inverse velocity) can be also calculated from the descendant level, say
 $E_{\zeta}$, as
 \cite{Otsu06b}
 \begin{equation}
  \zeta^{-1}=
   \lim_{L\to\infty}\frac{\Delta E_{\zeta}(L)}{2\pi/L}. 
   \label{eq_v}
 \end{equation}
 The corresponding excitation with small momentum can be found
 numerically.
 In calculating $c$ and $x$ from the excitation gaps via
 Eq.\ (\ref{eq_c_xs}),
 an estimate of $\zeta$ first needs to be obtained.
 However,
 it is not necessary in determining the BKT-transition points by
 Eqs.\ (\ref{eq_BKT_NEEL}) and (\ref{eq_BKT_DIMER}),
 because these are homogeneous equations of $x$, and thus the
 gaps---instead of dimensions---can be used.
 This is one of the advantages of the level-spectroscopy approach
 \cite{Otsu06b}.

 In the following, we shall provide our results from numerical
 calculations for systems up to size $L=20$.
 The methodological aspects of transfer-matrix calculations have been
 well explained in the literature
 \cite{Cast07}. 
 Furthermore,
 due to the sparse nature of the matrices, we can output all elements to
 hard disk.
 Then, using the 
 \texttt{ARPACK}
 library
 \cite{ARPACK},
 we can calculate the dominant eigenvalues of the nonsymmetric real
 matrices.

 \ZUsan

 As a demonstration we give the $K_d$ dependence of the scaling
 dimensions at $K_a=0$ in
 Fig.\ \ref{spector}(a),
 where $x_0$, $x_{1,2}$, and $x_3$ are plotted by
 solid,
 dotted,
 and
 dashed lines, respectively.
 While the data are for a system with $L=20$, we can see the excitation
 spectra approach quite close to the exact ones at the non-interacting
 point $K_d=0$
 \cite{Fish-Step63}.
 Furthermore,
 we find that the HC and the VC excitations interchange their behaviors
 at $K_d=0$, and the former (the latter) shows a level crossing with
 the doublet excitations at a certain negative (positive) value.
 According to theoretical predictions\
 (\ref{eq_BKT_NEEL}) and (\ref{eq_BKT_DIMER}),
 these can provide finite-size estimates of the BKT-transition
 points to the HC and the VC phases;
 we shall give some evidence to support our augment.
 In Fig.\ \ref{spector}(b), we exhibit extrapolations of the finite-size
 estimates to the thermodynamic limit according to Eq.\
 (\ref{eq_extrap}).
 The downward (upward) pointing triangles exhibit values of $-K_d$
 ($K_d$) at which the crossings between
  $x_0$ and $x_{1,2}$
 ($x_{1,2}$ and $x_3$)
 occur in the systems with $L=16$, 18, and 20.
 Their extrapolated values strongly agree with each other
 (i.e., their deviation is within {\discrepancy}),
 which is the obvious condition to be satisfied.
 In Figs.\ \ref{spector}(c) and \ref{spector}(d),
 we plot averaged values, i.e., the left-hand sides (LHSs) of
 Eqs.\ (\ref{eq_DLOG_NEEL}) and (\ref{eq_DLOG_DIMER}),
 as well as the dimensions at the BKT-transition points estimated in
 Fig.\ \ref{spector}(b).
 In both cases, the extrapolated values of the averages agree with the
 theoretical value of $\frac12$, which exhibits universal level
 splittings due to logarithmic corrections expected at the transition
 points.
 These observations show that both our strategy and numerical
 procedure are valid also for investigations of the IDM
 \cite{Otsu05Otsu08}.

 \ZUyon

 In Fig.\ \ref{phasediagram},
 we summarize our results of numerical calculations for the global phase
 diagram of our model\ (\ref{eq_Hamil}), in which the diagonal line
 (the center point) corresponds to the isotropic (noninteracting)
 system.
 Due to the fact that the phase diagram is symmetric about the line, it
 is sufficient to explicitly calculate one side of the whole parameter
 space; our calculations are thus restricted to the region $K_d\le0$
 (i.e., the upper-left triangular area).
 The open circles with solid lines give the phase boundaries between the
 dimer-liquid and the columnar phases.
 In the area apart from the BKT-transition boundaries, we can estimate
 the Gaussian coupling from the relation $K=\sqrt{{x_{1,2}}/{x_3}}$. 
 The contour lines of $K=\frac16$, $\frac14$, and $\frac12$, in addition
 to the points (plus marks) associated with $K=1$, $2$, and $3$ on the
 isotropic line are given in the figure.
 As expected, the dimer-liquid phase spreads over the area satisfying
 the condition $0<K\le4$.
 For $1\le K\le4$, this phase only survives on the isotropic line, but
 it is eventually terminated by another BKT transition caused by the
 marginally relevant ${\cal L}_4$ perturbation.
 We provide our estimation of the point by our approach (double circle
 in the figure) although some numerical results were previously
 available
 \cite{Alet05,Cast07}.
 We shall explain our method and compare our results with these in Sec.\
 \ref{sec_DISCUSSIONS}.

 \ZUgo

 To check the criticality of the dimer-liquid phase,
 we estimate the central charge along the line $K_h=0$ by the use of
 relations\ (\ref{eq_c_xs}) and (\ref{eq_v}).
 In Fig.\ \ref{eta-cv-c},
 we give the $K_v$ dependencies of
 the effective geometric factor $\zeta$ (diamonds),
 the coefficient of the $1/L$ correction $\gamma$ $(=c/\zeta)$ (squares), 
 and
 their product to estimate $c$ (circles).
 With increasing anisotropy, $\zeta$ deviates from the isotropic value
 of $1$ and approaches a certain value around {\anisotropy} in the limit
 $K_v\to-\infty$.
 Simultaneously, $\gamma$ declines in value and thus cannot itself give
 the universal amplitude of the finite-size correction.
 However, as expected, their product maintains a value $c=1$ within the
 dimer-liquid phase, and hence the proper normalization using the
 effective geometric factor is necessary for anisotropic systems.
 In the attractive region, one finds a point at which the central charge
 exhibits a steep decrease.
 We can check that the point is almost on the phase transition boundary
 to the VC phase (see the vertical arrow),
 and thus that it is consistent with the level-crossing calculations.

 The dimer-liquid region with $1\le K\le4$ corresponds to the unstable
 Gaussian fixed line; the ${\cal L}_2$ perturbation, except for $K=1$,
 brings about second-order phase transitions to the columnar phases.
 In this case, as $1/\xi\propto|K_d|^\nu$, the critical exponent
 characterizing the diverging correlation length is given by
 $1/\nu=2-X_{2,0}=2-2/K$
 \cite{Card86b}.
 To treat this transition, we have performed a finite-size-scaling
 analysis of the corresponding excitation gaps and have checked that
 the scaling behavior is very good although we do not provide the data
 here.
 In contrast,
 the liquid phase is absent in the more attractive region, whence the
 phase transition between the HC and the VC phases becomes first order
 (the solid line) accompanied by a jump in the phase-locking point
 $\langle\sqrt2\phi\rangle$ from 0 or $\pi$ to $\pi/2$ or $3\pi/2$.

 Finally, we discuss the transition to the strong repulsion phase
 (the upper-right gray-color region)
 at which the stiffness of the Gaussian model vanishes
 (see squares with solid lines in Fig.\ \ref{phasediagram}).
 In terms of the height model, this vanishing permits the interface to
 tilt globally without cost
 \cite{Alet05,Frad04}.
 To get some deeper insight, we shall here focus again on the analogy to
 a transition observed in the quantum-spin chain.
 The spin-$\frac12$ XXZ chain is solvable
 \cite{Cloi66,Yang66}
 and exhibits $c=1$ criticality for the anisotropy parameter satisfying
 $-1\le \Delta<1$.
 This phase is terminated at the SU(2) ferromagnetic point $\Delta=1$,
 where there occurs a first-order phase transition accompanied by the
 vanishing of the Gaussian coupling.
 At this point, the ground state of the $L$-site chain forms a SU(2)
 multiplet with total spin $L/2$ and thus possesses $L+1$ degeneracy
 with respect to the $z$ component $S^z_{\rm total}\in[-L/2,L/2]$.
 This degeneracy is also implied from the following theoretical
 observation:
 since the vertex operator corresponding to ${\cal O}_{0,n}$ in Sec.\
 \ref{sec_THEORY}
 expresses a $n$-spin flip excitation from the ground state with
 $S^z_{\rm total}=0$, the charge $n$ in a $L$-site system is restricted
 to values in $[-L/2,L/2]$.
 In addition, since the scaling dimension $X_{0,n}$ becomes zero
 in the limit of $K\to0$, at least the corresponding $L$ excited levels
 should degenerate to the ground state to realize the $L+1$ degeneracy
 \cite{Nomu96,Naka97}.
 Returning to our DM where the string number plays a role as a total
 magnetization, the ground-state energy in each topological sector,
 $E_{M,\rm g}(L)$, is expected to become independent of $M$.
 To see this degeneracy, we calculate the $M$-dependent free-energy
 density $Lf_M(L)=E_{M,\rm g}(L)$.
 In Fig.\ \ref{freeenergy}, we give the results for the isotropic system
 with $L=20$.
 The average of couplings varies within $-3\le K_a\le0$ and our estimate
 of the transition point is $K_a\simeq \KaFST$
 (see Sec.\ \ref{sec_DISCUSSIONS}).
 One can see that, with decreasing $K_a$, the free energy tends to
 show a weaker $M$ dependence and, for $K_a$ smaller than roughly the
 transition point, it becomes almost constant and zero.
 The above-mentioned $O(L)$ degeneracy inferred from the instability of
 the Gaussian criticality seems to be consistent with this $M$
 independence.
 However, there still exists a discrepancy in the degree of degeneracy
 with the staggered state; we shall discuss this issue for the rest of
 this section.

 \ZUnana

 It is known that the degeneracy of the staggered state is subextensive,
 i.e., $\propto\exp(aL)$
 \cite{Cast07,Bati04}.
 This is because, as depicted in Fig.\ \ref{dimerconf}(b), the $\pi/2$
 counterclockwise simultaneous rotation of all dimers along a dotted
 line in the $[11]$ direction can be performed {\it independently} of
 each other---the same holds also for clockwise rotations along the
 lines in the $[1\bar{1}]$ direction (a dashed line is an example).
 Thus, if one chooses
 a certain direction,
 the staggered state is completely ordered in that direction and
 completely disordered in the other one.
 While the problem of how the staggered state is stabilized is quite
 unclear, we shall try to give an insight based on an analysis of the
 one-string motion. 
 For convenience, we consider the transfer matrix connecting the
 next-nearest-neighbor rows, i.e., ${\bf T}_{1-L/2}^2$, so its
 eigenvalues or their logarithms are squared (i.e., $\lambda^2$) or
 doubled (i.e., $2E$), respectively.
 We treat two sites as one unit in which four states are included.
 Then, we can analytically diagonalize the matrix by a Fourier
 transformation and obtain the $q$-dependence of the energy $2E_q$,
 i.e., the ``dispersion relation'' of the one-string motion in the $x$
 direction.
 Here, we only show results; the details of how to construct the
 transfer matrix and also the calculation of eigenvalues in the
 one-string sector are given in the Appendix.
 In Fig.\ \ref{oneparticle}, for several values of the interactions
 (isotropic cases), we draw the lower two of four bands.
 When the eigenvalue becomes a complex number, we take its magnitude in
 the plot.
 While the symmetric two-band structure for the non-interacting case
 [Fig.\ \ref{oneparticle}(a)]
 \cite{Kast61,Alet05}
 is deformed by interactions, there is a
 unique minimum at $q=0$ for finite interaction cases
 [Figs.\ \ref{oneparticle}(b) and \ref{oneparticle}(c)].
 At the same time, as expressed by the dotted lines, complex-conjugate
 pairs of eigenvalues start to appear near the zone boundary points
 $q=\pm\pi$.
 And then,
 in the strong repulsion limit [Fig.\ \ref{oneparticle}(d)], we find an
 emergence of a two-fold degenerate zero-energy flat band.
 The corresponding eigenvalues are given by
 $\lambda_q^2={\rm e}^{\pm iq}$,
 which represents the modulations with wave number $\pm q$ in the $y$
 direction
 \cite{Liebmann}.
 Consequently,
 our results show dispersionless motion in the $[11]$ and the
 $[1\bar{1}]$ directions, which precisely reflect the above-mentioned
 degeneracy, and thus this flat band structure may be a signature of the
 subextensive degeneracy in the staggered state.
 If we accept this naive argument, we can conjecture that the staggered
 state is only realized in the limit.
 However, our argument is of course at a very speculative level; a full
 understanding should include also the degeneracy in many-string
 sectors.

 \ZUhachi

 \section{DISCUSSIONS and SUMMARY}
 \label{sec_DISCUSSIONS}

 The isotropic case was discussed in detail in Refs.\
 \cite{Alet05,Cast07},
 where the BKT-transition point driven by ${\cal L}_4$ and the
 first-order transition point to the strong repulsion phase were
 numerically obtained.
 We have also estimated these transition points (see the double circle
 and the double square in Fig.\ \ref{phasediagram}); in particular, for
 the former, the above level-spectroscopy approach has been applied.
 Thus, here we briefly explain our procedure and compare the results.
 Since the Lagrangian density ${\cal L}_0+{\cal L}_4$ is analyzed in the
 region $K\simeq4$, we focus our attention not on $O_{0,3}$ but instead
 on the following order parameters responsible for the breaking of the
 $\pi/2$ rotational symmetry:
 \begin{align}
  O_4&=\sqrt2\cos2\sqrt2\phi,
  \label{eq_2Neel}\\
  O_5&=\sqrt2\sin2\sqrt2\phi.
  \label{eq_2Dimer}
 \end{align}
 While the former is the orientational order parameter, the latter
 represents the plaquette order
 \cite{Ralk08}
 which has not been found in classical DMs
 \cite{RK,Cast07}
 (the locking points are
 $\langle\sqrt2\phi\rangle=\pi/4$, $3\pi/4$, $5\pi/4$, and $7\pi/4$).
 One then finds that via the transformation
 $2\sqrt2\phi\to\sqrt2\phi$ and $K/4\to K$,
 the Lagrangian density and the operators are reduced to
 \begin{align}
  {\cal L}_0+{\cal L}_4(y_4)\to{\cal L}_0+{\cal L}_2(y_4),~~
  O_{4,5} \to O_{0,3}.
 \end{align}
 Therefore, from the discussion in Sec.\ \ref{sec_THEORY},
 the level-crossing and the level-splitting conditions\
 (\ref{eq_BKT_NEEL}) and (\ref{eq_DLOG_NEEL})
 are satisfied by the scaling dimensions of these operators,
 say $x_{4,5}$
 (here, we have taken the condition $y_4<0$ into account).
 Since the half-charge excitations $\exp(\pm i\frac12\sqrt2\theta)$ are
 absent in our system, we employ the condition
 \begin{align}
  \frac{3x_{4}+ x_5}{4}=\frac12
  \label{eq_LS_L4}
 \end{align}
 to determine the BKT-transition point.
 The corresponding excitation levels can be found in the sectors
 specified by their symmetries given in Table\ \ref{tabI}.
 We extrapolate the finite-size estimates up to $L=20$ to the
 thermodynamic limit according to Eq.\ (\ref{eq_extrap}).
 We then obtain the BKT-transition point as $K_a\simeq\KaBKT$.
 As we see in Table\ \ref{tabII}, the agreement with previous results is
 very good, which indicates that our approach is valid.
 Similarly, for first-order phase transition, we have estimated the
 point via condition $K=0$ (see also Ref.\ \cite{Cast07}) while
 others have determined this transition from a point of breakdown in the
 condition $c=1$.
 Our result
 is closer to
 the estimate of Alet et al.
 \cite{Alet05}
 although there still exists considerable discrepancy among these
 estimates.
 Likewise for instance for phase-separation transitions observed in
 one-dimensional electron systems, higher-order corrections have been
 argued to ambiguously affect estimations
 \cite{Naka97}.
 Thus, we think that the discrepancy in these estimates may reflect
 their effects.

 To summarize, we investigated the anisotropically interacting dimer
 model on a square lattice.
 For the attractive case, the orientational-order-parameter perturbation
 introduced by the anisotropy brings about the BKT transition to the
 columnar phases.
 We pointed out the close relationship of our model to a frustrated
 quantum-spin chain and then found the criteria to determine the
 transition points.
 Using these, we performed level-spectroscopy analysis of the eigenvalue
 structures of the transfer matrices.
 Numerical results were then summarized as the global phase diagram
 (Fig.\ \ref{phasediagram}), which includes the dimer-liquid, the
 columnar, and the strong repulsion phases.
 Furthermore, we checked the level-splitting conditions and evaluated
 the value of the central charge, which provided solid evidence to
 confirm the universality of the phase transition.
 By contrast, for the repulsive case, although we determined the
 dimer-liquid phase boundary, there exist some points with unclear
 status within the strong repulsion phase including the staggered state.
 Based on the dispersion relation of the one-string motion, we gave a
 possible scenario for the stabilization of the staggered phase.
 However, although this issue still remains an open question, we now
 think that the nature of the nonsymmetric real matrix might have
 relevance to its description
 \cite{Liebmann}.

 \HYOUni

 \section*{ACKNOWLEDGMENT}

 The author thanks
 Y. Tanaka,
 M. Fujimoto,
 K. Kobayashi, 
 and 
 K. Nomura
 for stimulating discussions. 
 Most of the computations were performed using the facilities of
 Information Synergy Center in Tohoku University. 
 This work was supported by
 Grants-in-Aid from the Japan Society for the Promotion of Science,
 Scientific Research (C), Contract No.\ 17540360.

 \appendix

 \ZUqu

 \section{A TRANSFER-MATRIX CALCULATION IN ONE-STRING SECTOR}

 In this appendix,
 we shall explain how to construct the transfer matrix in the one-string
 sector and an analytical calculation of eigenvalues by the use of a
 Fourier transformation.
 Since a row of the reference configuration given in Fig.\
 \ref{dimerconf}(b)
 expresses a string vacuum state, we write it as $\left|0\right\rangle$
 (see the top left in Fig.\ \ref{states-elements}).
 For convenience,
 we treat two sites in the $x$ direction as one unit cell which includes
 four bonds (see squares by dotted blue lines).
 Then,
 one-string states can be obtained via replacements of one of $L_c$
 $(=L/2)$ unit cells in $\left|0\right\rangle$ by four possible dimer
 configurations.
 From left to right of the first line in Fig.\ \ref{states-elements}
 (except for the vacuum),
 we call these as
 $\left|A_l\right\rangle$,
 $\left|B_l\right\rangle$,
 $\left|C_l\right\rangle$, and
 $\left|D_l\right\rangle$, respectively. 
 Here, the center is supposed to be an $l$th unit cell ($l\in[1,L_c]$).
 Now, consider transfers of one-string states to those in the
 next-nearest-neighbor row in the $y$ direction.
 Then, one can find 24 microscopic processes, which are listed in
 subsequent lines in Fig.\ \ref{states-elements}.
 For instance, the second line shows seven microscopic processes of
 transfers from $\left|A_l\right\rangle$ in the first row to states
 in the third row, and thus exhibits an operation of the transfer
 matrix, i.e., ${\bf T}_{1-L_c}^2\left|A_l\right\rangle$.
 Consequently, we can obtain the following recursion relations for the
 transfers in the one-string sector
 \begin{widetext}
  \begin{align}
   {\bf T}_{1-L_c}^2
   \left|A_l\right\rangle
   =&    \left|C_{l-1}\right\rangle
   + v^2 \left|A_l    \right\rangle
   + h   \left|B_l    \right\rangle
   + h   \left|C_l    \right\rangle
   +     \left|D_l    \right\rangle
   +     \left|B_{l+1}\right\rangle
   +     \left|D_{l+1}\right\rangle,\\
   {\bf T}_{1-L_c}^2
   \left|B_l\right\rangle
   =&h   \left|C_{l-1}\right\rangle
   + v^2 \left|A_l    \right\rangle
   + h^2 \left|B_l    \right\rangle
   + h   \left|C_l    \right\rangle
   + h   \left|D_l    \right\rangle
   +     \left|B_{l+1}\right\rangle
   +     \left|D_{l+1}\right\rangle,\\
   {\bf T}_{1-L_c}^2
   \left|C_l\right\rangle
   =&    \left|C_{l-1}\right\rangle
   + v^2 \left|A_l    \right\rangle
   + h   \left|B_l    \right\rangle
   + h^2 \left|C_l    \right\rangle
   +     \left|D_l    \right\rangle
   + h   \left|B_{l+1}\right\rangle
  + h   \left|D_{l+1}\right\rangle,\\
   {\bf T}_{1-L_c}^2
   \left|D_l\right\rangle
   =&v^2 \left|C_{l-1}\right\rangle
   + v^2 \left|B_l    \right\rangle
   + v^2 \left|D_l    \right\rangle, 
  \end{align}
 \end{widetext}
 where coefficients represent the Boltzmann weights of interactions
 in the first and the second rows.
 Next, by the use of the Fourier transformation,
 we can block-diagonalize the representation of ${\bf T}_{1-L_c}^2$:
 suppose that
 \begin{align}
  \left|X_q\right\rangle
  =
  \frac{1}{\sqrt{L_c}}\sum_{l=1}^{L_c}e^{-iql} \left|X_l\right\rangle~~~
  (X=A, B, C, D),
 \end{align}
 then the $q$-block representation spanned by states
 $\{\left|A_q\right\rangle$,
 $\left|B_q\right\rangle$,
 $\left|C_q\right\rangle$,
 $\left|D_q\right\rangle\}$
 is given by a $4\times4$ complex nonsymmetric matrix:
 \begin{equation}
  [{\bf T}_{1-L_c}^2]_q
   =
   \begin{pmatrix}
    v^2 & h  +  z_q & h  + \bar z_q  & 1+  z_q \\
    v^2 & h^2+  z_q & h  +h\bar z_q  & h+  z_q \\
    v^2 & h  +h z_q & h^2+ \bar z_q  & 1+h z_q \\
    0   & v^2       & v^2  \bar z_q  & v^2     \\
   \end{pmatrix},
 \end{equation}
 where $z_q,\bar z_q\equiv e^{\pm iq}$. 
 Hence, the characteristic equation to determine eigenvalues $\rho$ is
 given by 
 \begin{widetext}
  \begin{align}
   \rho^4
   -2 (\cos q+v^2+h^2) \rho^3
   +[v^4-4h(1-h)v^2+(1-h^2)^2] \rho^2
   +2v^2(1-h)^2(1-v^2-h^2) \rho
   +v^4(1-h)^4=0. 
   \label{characteristicequation}
  \end{align}
 \end{widetext}
 Since it is invariant under a transformation $q\to-q$, a $q$ dependence
 of the eigenvalue structure is even with respect to the point $q=0$.
 Meanwhile, in general cases we use a software to evaluate
 $q$ dependences of eigenvalues; in some limiting cases,
 Eq.\ (\ref{characteristicequation}) becomes simple and permits us to
 easily manipulate:
 for instance, for the noninteracting case $h=v=1$, two of the four
 eigenvalues are zero, and the rest is obtained from an equation
 $\rho^2 -2(\cos q+2)\rho +1=0$.
 It then provides two real bands, as given in Fig.\
 \ref{oneparticle}(a).
 In contrast, for the strong repulsion limit $h=v=0$,
 two of the four eigenvalues are zero again,
 but others are complex values with a modulus of 1,
 i.e., $e^{\pm iq}$
 [see Fig.\ \ref{oneparticle}(d)].
 An implication of this eigenvalue structure, in particular a
 correspondence to the degeneracy of states in the IDM is discussed in
 the last part of Sec.\ \ref{sec_NUMERICAL}.

 \newcommand{\AxS}[1]{#1,}
 \newcommand{\AxD}[2]{#1 and #2,}
 \newcommand{\AxT}[3]{#1, #2, and #3,}
 \newcommand{\AxQ}[4]{#1, #2, #3, and #4,}
 \newcommand{\AxP}[5]{#1, #2, #3, #4, and #5,}
 \newcommand{\AxH}[6]{#1, #2, #3, #4, #5, and #6,}
 \newcommand{\REF }[4]{#1 {\bf #2}, #3 (#4)}
 \newcommand{\JPSJ}[3]{\REF{J. Phys. Soc. Jpn.\  }{#1}{#2}{#3}}
 \newcommand{\PRL }[3]{\REF{Phys. Rev. Lett.\    }{#1}{#2}{#3}}
 \newcommand{\PRA }[3]{\REF{Phys. Rev.\         A}{#1}{#2}{#3}}
 \newcommand{\PRB }[3]{\REF{Phys. Rev.\         B}{#1}{#2}{#3}}
 \newcommand{\PRE }[3]{\REF{Phys. Rev.\         E}{#1}{#2}{#3}}
 \newcommand{\NPB }[3]{\REF{Nucl. Phys.\        B}{#1}{#2}{#3}}
 \newcommand{\JPA }[3]{\REF{J. Phys.\ A:         }{#1}{#2}{#3}}
 \newcommand{\JPC }[3]{\REF{J. Phys.\ C:         }{#1}{#2}{#3}}
 \newcommand{\IBID}[3]{\REF{{\it ibid.}}{#1}{#2}{#3}}

 \end{document}